# THE WESTERN AUSTRALIAN OPTICAL GROUND STATION

*Shane Walsh[1*], Alex Frost[1], William Anderson[1], Toby Digney[1], Benjamin Dix-Matthews[1,2], David Gozzard[1,2], Charles Gravestock[1], Lewis Howard[1], Skevos Karpathakis[1], Ayden McCann[1], Sascha Schediwy[1,2]*

[1]*International Centre for Radio Astronomy Research, The University Western Australia, Perth, Australia*
[2]*Australian Research Council Centre of Excellence for Engineered Quantum Systems, Department of Physics, The University of Western Australia, Perth, Australia*
*\*shane.walsh@uwa.edu.au*

## Abstract

Free-space communications at optical wavelengths offers the potential for orders-of-magnitude improvement in data rates over conventional radio wavelengths, and this will be needed to meet the demand of future space-to-ground applications. Supporting this new paradigm necessitates a global network of optical ground stations. This paper describes the architecture and commissioning of the Western Australian Optical Ground Station, to be installed on the roof of the physics building at the University of Western Australia. This ground station will incorporate amplitude- and phase-stabilisation technology, previously demonstrated over horizontal free-space links, into the ground station's optical telescope. Trialling this advanced amplitude- and phase-stabilisation technology, the ground station will overcome turbulence-induced noise to establish stable, coherent free-space links between ground-to-air and ground-to-space. These links will enable significant advances in high-speed and quantum-secured communications; positioning, navigation, and timing; and fundamental physics.

## 1. Introduction

The demand for data is ever increasing [1], with the advent of small satellites dramatically increasing the number of spacecraft in orbit, each with increasing capacity to generate data. Spacecraft provide global communications and broadcast coverage, bring broadband internet to regions where geography or lack of economic viability preclude terrestrial connections, and provide orbital platforms for weather and disaster monitoring, military and civilian surveillance, as well as a host of scientific endeavours. Moving all that data requires high bandwidth links, and traditional radio frequency (RF) links are already causing a bottleneck in free-space communications.

This bottleneck is even more of a concern beyond Earth orbit. Despite being the most significant television broadcast in history, with the resource investment that goes along with such an event, the Apollo 11 Moon landing was transmitted at a quality well below even 1969 standards due to the limited ~50 kb/s data rate possible from the spacecraft S-Band RF communications system [2]. NASA's Artemis II mission aims to return humans to Moon as soon as 2023, and to stream this landing at a modern 4K resolution would require a bandwidth of ~50 Mb/s, a thousand times higher than Apollo 11. This is before even considering the bandwidth needed for telemetry, tracking and control. To meet the hundreds of Mb/s data link requirements of future spacecraft, particularly crewed missions [3], more than a thousand-fold gain in bandwidth is needed.

The data rate achievable via free space propagation is proportional to the transmitter power, transmitter and receiver antenna areas, and the square of the carrier frequency. Increasing the transmitter power or antenna size by an order of magnitude on a spacecraft is not practical, nor is an order magnitude increase in ground station antenna area. But if the carrier frequency is moved from radio (GHz) to optical (hundreds of THz), significantly higher data rates become feasible [4]. Optical frequencies have already revolutionised terrestrial communications via optical fibre [5], and offer the same prospect for free-space communications. Optical ground stations are identified as a priority in the Australian Space Agency's Communications Technologies and Services Roadmap 2021-2030 [6].

NASA's first attempt at a bidirectional free-space optical (FSO) link between the ground and space occurred in 2013 with the Lunar Laser Communication Demonstration (LLCD): This link achieved record data rates of 20 Mb/s for uplink and 622 Mb/s for downlink between the Lunar Atmosphere and Dust Environment Explorer (LADEE) in lunar orbit and receiving stations on the ground [7]. These same ground stations, OGS-1 located in California and OGS-2 in Hawaii, will support the Laser Communications Relay Demonstration (LCRD), a longer-term mission launching in 2021 to demonstrate a bidirectional 1.244 Gb/s optical link between the ground and Geosynchronous Earth Orbit (GEO) [8]. To demonstrate optical communications from deep space, the Psyche spacecraft will carry the Deep Space Optical Communications (DSOC) experiment on its mission to the main belt asteroid 16 Psyche. Launching in 2022, DSOC is expected to achieve a data rate of 1.2 Mb/s from a distance of 2.62 AU (390 million km), comparable to the maximum distance between Earth and Mars [9].

Demonstration of optical communications as an operational link is planned for the Artemis II mission. The Optical to Orion (O2O) project aims to provide the Orion Multi-Purpose Crew Vehicle with a bidirectional optical link with data rates of up to 250 Mbps for downlink and 20 Mbps for uplink [3], sufficient to live-stream the first woman and next man on the moon in 4K resolution.

In addition to substantially improved data rates, free-space optical communication links have other advantages over traditional RF links: The higher directionality provides more efficient power transmission, lower probability of interception, and in most countries, an unregulated spectrum. However, fully capitalising on these advantages for free-space communications through the atmosphere requires overcoming beam fluctuations caused by atmospheric turbulence: Longitudinal fluctuations vary the time-of-flight of optical signals, thereby increasing the phase noise, while transverse fluctuations produce beam wander and scintillation, resulting in deep fades and additional amplitude noise [10]. Amplitude-stabilisation is an established practice in optical astronomy through the use of adaptive optics [11, 12], and phase-stabilisation is being demonstrated by various groups over horizontal links [13-15] and links to airborne targets [16]. Equipping ground stations with these capabilities will unlock the full bandwidth potential of free-space optical links.

As turbulence mitigation techniques gain maturity and FSO communications move from demonstrator missions to primary operational links, a global network of ground stations is needed for global coverage and weather redundancy [17]. Western Australia offers an ideal location with large, unpopulated areas devoid of light pollution, large fraction of cloudless days, and unique sky coverage [18]. With this impetus, we are commissioning the Western Australian Optical Ground Station (WAOGS), benefiting from favourable geography, and uniquely poised to exploit our group's atmospheric amplitude- and phase-stabilisation technology. Capable of supporting multiple optical communication schemes, WAOGS will form one node of the Australian Optical Ground Station Network, with two further nodes being developed by Defence Science and Technology Group (South Australia) and the Australian National University (Australian Capital Territory) [19].

In this paper, we describe the current status of WAOGS and outline the path forward from technology demonstrator to operational ground station. Section 2 briefly describes the challenge of FSO through the atmosphere. Section 3 provides an overview of the facility in its current commissioning phase and describes each subsystem. Section 4 presents current work in progress and commissioning results to date. Section 5 identifies the work required to transition from the commissioning phase to ground-to-space operations. Sections 6 and 7 provide a discussion and concluding remarks.

## 2. Atmospheric FSO Communications

While optical communication through fibre has become the backbone of the internet and ground connectivity, radio frequencies have remained the standard for wireless communications. This is largely in part due to the atmospheric effects that have limited impact on radio signals but present a significant challenge for optical signals. While effects such as beam spread and transmission losses are of concern, discussion here is limited to the particular effects of atmospheric turbulence that require real-time correction: Amplitude noise caused by beam wander and scintillation, and phase noise from time-of-flight variations.

The atmosphere is constantly in motion, with thermal energy being deposited and transferred through to smaller and smaller spatial scales. This energy transfer manifests as temperature variations, with the accompanying variation in refractive index along paths of any practical length. The particular effect this has on a FSO link depends on the integrated turbulence strength and the beam size ($D$) through the path. Turbulence strength is characterised by the refractive index structure constant $C_n^2$, which can be integrated over the path to give the Fried coherence length $r_0$. This parameter is the diameter of a patch of wavefront that remains coherent, defined as rms phase error of less than 1 radian. It is wavelength dependent and is typically of the order of a few centimetres for a very turbulent path (such as a horizontal link over kilometres) to a few tens of centimetres over relatively stable paths (vertical links from favourable sites such as astronomical observatories).

For weaker turbulence, or a small beam size ($D/r_0 < 1$), the variations in refractive index as the beam propagates cause small angular deviations from the initial propagation axis. The beam arriving at the receiver maintains coherence but has deviated from the optimum pointing and has experienced beam wander. This can cause the highest intensity region of the beam to fall off the receiver, or depending on beam and receiver geometry, cause the beam to miss the receiver entirely. Either way, the result is a reduction in fibre coupling and received power.

If turbulence is strong, or for a large beam size ($D/r_0 > 1$), the same angular deviations occur but different regions of the beam cross-section undergo different deflections. The beam arriving at the receiver will consist of individually coherent patches of size $\sim r_0$, each with an uncorrelated angle of arrival. As with beam wander, this causes a reduction in fibre coupling, but it also has the additional issue of these regions randomly constructively and destructively interfering with each other. This scintillation produces strong intensity fluctuations above and beyond what occurs with beam wander alone and occurs even when the overall beam pointing remains on target.

The same refractive index variations that cause angular deviations also cause temporal time-of-flight variations, introducing phase noise. The amplitude noise from beam wander and scintillation is a problem for any FSO link through the atmosphere (direct detection or coherent), but phase noise is an added problem for coherent links. Direct detection links are limited to amplitude modulation schemes such as On/Off Keying (OOK) and Pulse Position Modulation (PPM), while coherent links have the added degrees of freedom of phase and/or frequency modulation. Coherent signals are the standard for radio frequency links and high bandwidth optical fibre links, and these extra degrees of freedom enable higher data rates, improved background rejection, and more sensitive receivers than direct detection methods. A phase-stabilised link also allows novel communication techniques such as Continuous Variable Quantum Key Distribution [20] and Twin Field Quantum Key Distribution [21, 22].

Turbulence varies on timescales of tens of milliseconds, and therefore requires methods of sensing and correcting the amplitude and phase noise that can operate faster than these timescales. Amplitude-stabilisation is a mature technology, originally conceived for astronomy in the 1950s [23], developed by the US military for defence purposes in the following decades, and picked up again by the astronomical community after declassification and development of sufficient computing power in the 1990s. Without this adaptive optics (AO) technology, atmospheric turbulence renders the resolution of a 10 m telescope no better than that of a 10 cm telescope. With adaptive optics to counter beam wander and higher order aberrations in real-time, telescopes can approach their diffraction limit for imaging, or as needed in this context, reduce deep-fades and scintillation to maintain fibre coupling of optical signals.

Phase-stabilisation is a more nascent technology. Through interferometric phase measurement of the round-trip signal, a servo loop applies a frequency shift to advance and delay the signal to counter the phase noise. Phase stabilisation is being developed for metrology applications but will improve the sensitivity of coherent detection methods (which rely on a phase locked local oscillator), improving the bit error rate (BER) of coherent modulation techniques such as quadrature phase shift keying (QPSK).

In summary, amplitude-stabilisation is needed to counter beam wander and scintillation to maintain power to the receiver fibre, and phase-stabilisation is needed to maintain the phase of the local oscillator with the signal for coherent detection. These are the technologies that WAOGS will be commissioning for FSO links.

## 3. WAOGS Phase I: Commissioning

The Western Australian Optical Ground Station is built around a PlaneWave Instruments CDK700 observatory grade telescope. Installed for initial commissioning within the Ken and Julie Michael Building at the University of Western Australia (Figure 1), the telescope is being prepared for relocation to the roof of the Physics building for commencement of on-sky commissioning.

The telescope serves as receiver and transmitter for optical signals (presently C-band centred at 1550 nm), feeding through one of the Nasmyth ports to the amplitude-stabilisation system mounted directly to the telescope. This amplitude-stabilisation counters turbulence-induced beam wander to reduce deep fades and maintain fibre coupling. The amplitude-stabilised signal can then be fed through fibre to a phase-stabilisation system, located in a thermally controlled environment.

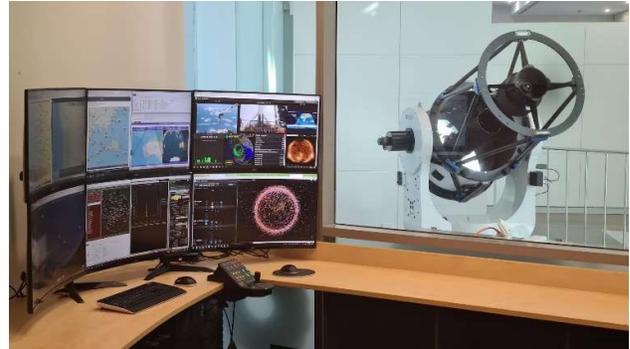

Fig. 1: The WAOGS commissioning control room and telescope.

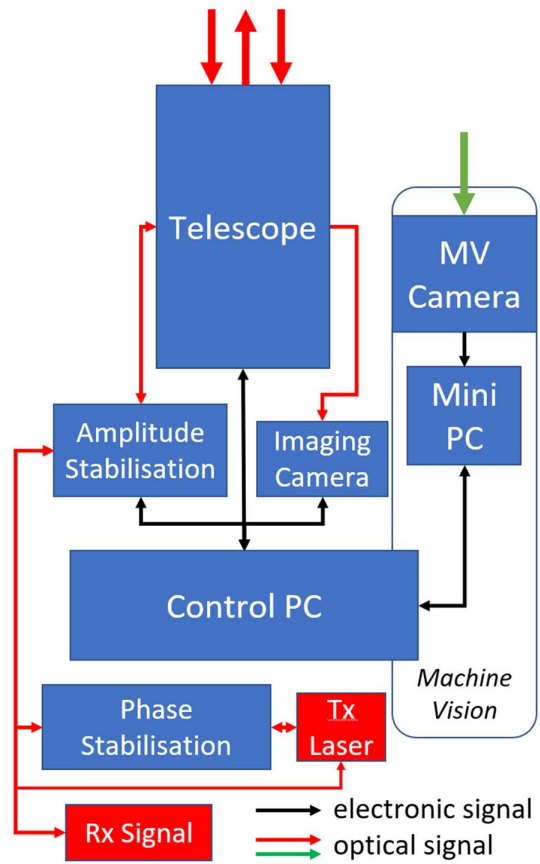

Fig. 2: Block diagram of WAOGS architecture.

The ground station will acquire an airborne or spaceborne target via GPS or TLE orbit propagation. Machine vision will then fine point the telescope to place the target's signal on a quad photodetector (QPD). The QPD error signals will drive piezo-actuated tip/tilt mirrors (TTM) to correct the pointing of both the received and transmitted signals for amplitude stabilisation. A system diagram is presented in Figure 2, with each system detailed in the following subsections.

### 3.1. Drone:

For vertical link commissioning we are employing a DJI Matrice 600 Pro professional grade drone (Figure 3, top). The drone has a maximum horizontal speed of 65 km/h, maximum altitude of 2500 m, and an unencumbered hovering time of 32 minutes. It can maintain hover in wind speeds up to 8 ms$^{-1}$.

The drone will carry a payload consisting of a camera for ground station acquisition and four 525 nm beacon LEDs for machine vision tracking circumscribing a 2-inch retroreflector for beam return (Figure 3, bottom). A GPS transmitter attached to the landing gear will transmit coordinates via LoRa to the ground station for coarse telescope pointing. Drone flight paths will be chosen to replicate the apparent motion of spacecraft ranging from LEO to cis-lunar space while links are established from the ground station. The drone will maintain payload orientation via the payload camera.

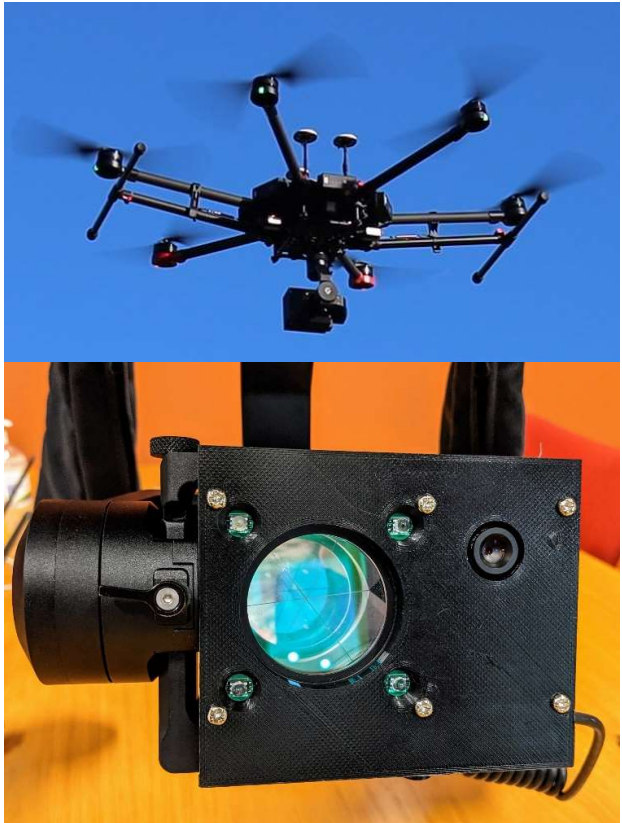

Fig. 3: *Top* – The DJI Matrice 600 Pro drone in flight, showing payload suspended below. *Bottom* – The drone optical payload.

### 3.2. Telescope:

The PlaneWave Instruments CDK700 is an observatory grade telescope (Figure 1) employing a corrected Dall-Kirham (CDK) optical design, featuring a 0.7 m elliptical primary mirror and spherical secondary focusing an f/6.5 beam to two Nasmyth ports, selectable by a rotating flat tertiary mirror. The telescope forms one element of a Keplerian beam expander with the other end of the beam expander residing within the amplitude-stabilisation optics mounted to one of the Nasmyth ports. For commissioning, the system is configured as a 15× beam expander, launching a 34 mm beam from a sub-aperture of the telescope. This configuration assures that the beam size is smaller than $r_0$, meaning first order tip/tilt is sufficient for amplitude stabilisation. It also prevents the 2-inch retroreflector on board the drone from clipping the beam, maximising the power returned to the telescope.

The telescope is altitude-azimuth mounted with a pointing precision of 2 arcseconds (~10 µrad), and capable of open-loop sidereal tracking with less than 1 arcsecond (~5 µrad) rms error over 10 minutes with a suitable pointing model. It will be installed in a 3.5 m diameter motorised dome on the roof of the physics building at the UWA Crawley campus with remote operation capabilities. Telescope control is implemented through an HTTP API with a custom GUI incorporating the telescope and other systems.

The remaining Nasmyth port will house an imaging camera. In the context of ground station operations, it is needed for telescope pointing models and site characterisation. It can be made available for other purposes such as space situational awareness, astronomy, and outreach/training when the ground station might otherwise remain idle.

### 3.3. Machine Vision:

While the telescope is capable of pointing with an accuracy of 2 arcseconds (~10 µrad), it requires the target position to be known to that accuracy. The drone is equipped with GPS that can provide a location to meter-level accuracy, with the greatest uncertainty in altitude. This is sufficient for coarse telescope pointing, but finer pointing is needed to align the transmitted beam with the drone's retroreflector and place the returned beam on the QPD. To achieve this, we optically track the target with a machine vision system.

Mounted to the top side of the telescope tube, an f = 100 mm machine vision lens images to a Teledyne FLIR Blackfly S BFS-U3-28S5M-C camera, providing a 5.0° × 3.8° field of view on the 8.8 mm × 6.6 mm sensor. This configuration balances the need for a sufficiently large field of view to capture the drone within GPS pointing accuracy (a few degrees) with a pixel scale fine enough to guide the transmitted beam to the retroreflector and the returned beam onto the QPD. A filter centred at 525 nm with 92 nm bandwidth is used to reduce background light and improve the signal to noise ratio of the 525 nm beacon LEDs of the drone optical payload.

The machine vision control and processing are performed by an Nvidia Jetson Nano mini PC, connected to the camera via USB, also mounted to the telescope. The Jetson itself interfaces remotely over Ethernet from the Control PC. When commanded from the control PC, the Jetson starts execution of its control loop. This control loop initiates continuous image acquisition from the camera, analyses each image to detect the drone's beacon LEDs, and streams the pixel coordinates of the target via Ethernet to the control PC. The pixel coordinates are determined from the weighted centre of mass of all pixels above a threshold level.

The control PC converts the $x$ and $y$ pixel coordinates of the target to altitude and azimuth offsets $\Delta alt$ and $\Delta az$. A proportional and integral controller takes these offsets and calculates the offset *rate* to add to the telescope tracking to minimize these offset errors and maintain the target lock. Offset rates are used rather than offsets because they are less sensitive to the asynchronicity between machine vision measurements and the telescope's execution of commands. They are also more robust to interruption; if a frame is dropped or packet lost, the telescope will continue tracking at the last calculated rate rather than pause until a new command is issued.

### 3.4. Amplitude Stabilisation:

Amplitude stabilisation is achieved via a conventional tip/tilt adaptive optics system correcting both the uplink and downlink signals [24]. Due to the reciprocity of atmospheric turbulence, the correction of the received beam tip/tilt is also applicable to the transmitted beam when the return path is the same. This is valid for drone commissioning, but as discussed in Section 4, not necessarily valid for ground-to-space links.

Figure 4 illustrates the optical path of the amplitude-stabilisation system. The transmitted beam arrives via fibre to the amplitude-stabilisation system where a fibre-to-free-space collimator produces a 2.27 mm collimated beam. A 50:50 beamsplitter (needed for the return beam) reflects 50% of the beam to a beam dump to avoid undesired reflections, while the remaining beam is transmitted to a series of two tip/tilt mirrors (as well as other fold mirrors to contain the optical path within the breadboard). The collimated beam is then focused by an f = 300 mm lens, with its focus coincident with that of the telescope (f = 4540 mm) forming a ~15× Keplerian beam expander. The beam exits the telescope as a collimated beam of 34 mm diameter.

The transmitted beam propagates through the atmosphere until it is retroreflected by the drone. The reflected beam will return through the system along the same path as the transmitted beam, until it arrives at the beamsplitter. Now 50% of the return signal is transmitted back through the fibre, while 50% is reflected to the QPD assembly. This includes a 5× Galilean beam expander to optimise spot size and an f = 200 mm lens to focus the beam onto the QPD. The QPD measures angular deviation of the return beam (caused by atmospheric turbulence, but also pointing errors from mechanical misalignment, vibration etc.) as centroid shifts, which is the error signal used to drive the tip/tilt mirrors: the mirrors' tip and tilt angles are controlled to maintain the beam spot centred on the QPD, thereby maintaining the transmitted beam centred on the drone retroreflector, and the return beam centred on the fibre.

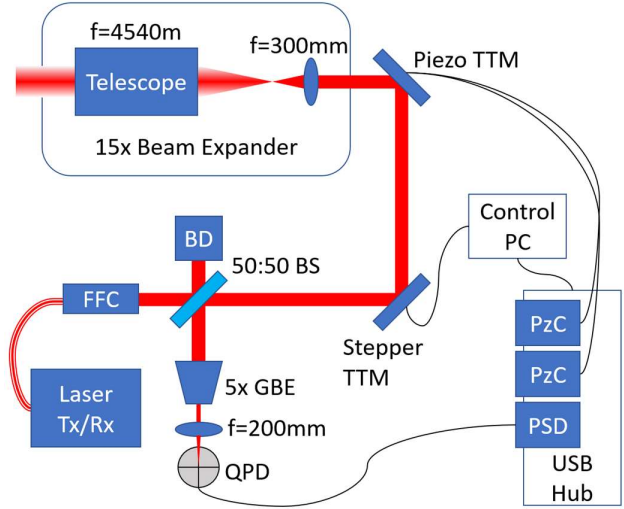

Fig 4: Block diagram of amplitude stabilisation system. *FFC:* fibre-to-free space collimator, *QPD:* Quad photodetector, *GBE:* Galilean beam expander, *BD:* Beam dump, *TTM:* Tip/tilt mirror, *PzC:* Piezo controller, *PSD:* Position sensing detector. Static fold mirrors not shown for simplicity. Red denotes an optical signal, while black is electronic.

Two tip/tilt mirrors are required to have the actuation range needed to counter the angular deflections expected from turbulence (a few µrad). Since there is a 15× beam expansion between the TTMs and the free-space path, their on-path actuation range is reduced by that same factor. The stepper TTM will correct the lower frequency, but higher amplitude fluctuations while piezo-actuated TTM will correct the higher frequency, lower amplitude fluctuations. Figure 5 depicts the amplitude-stabilisation system mounted to the telescope.

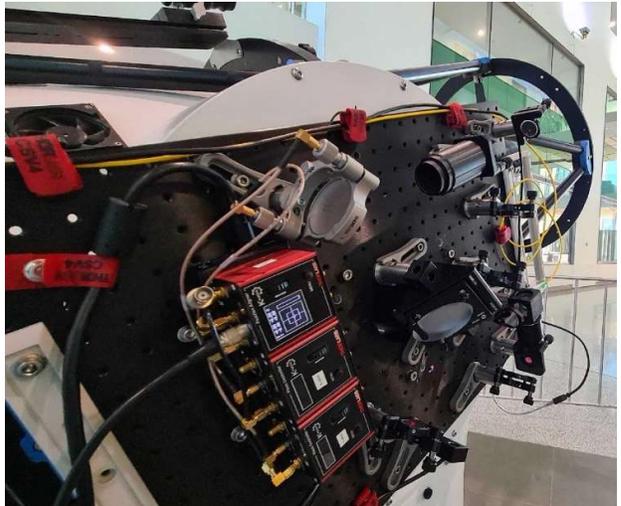

Fig. 5: The amplitude-stabilisation optics mounted to the telescope's Nasmyth port.

## 3.5. Phase Stabilisation:

Phase stabilisation is achieved via interferometric phase measurement of the round-trip signal [13, 24, 25]. The signal is split with the free-space link forming the long arm of an imbalanced Michaelson interferometer and the remaining signal forming the reference arm. The laser signal is sent to a splitter, diverting 1-2% to a Faraday mirror, and the remainder through an acousto-optic modulator (AOM). The AOM applies a nominal frequency shift to the signal, which is to be varied by the control loop. The frequency shifted transmitted beam is sent through the amplitude-stabilisation system and the telescope, propagating through the turbulent atmosphere where time of flight variations impart phase noise to the signal. The beam is reflected by the drone, where due to reciprocity, the phase noise is doubled. After returning through the telescope and amplitude-stabilisation system, the reflected beam is split between the transmitter and receiver assemblies. The portion that returns through the transmitter interferes with the reference arm, forming a heterodyne beat. The frequency of this beat is a function of the phase noise across the link, which can be used to generate an error signal to control the transmitter AOM. The AOM adjusts the frequency of the transmitted signal to maintain the phase of the transmitted signal with that of the return signal.

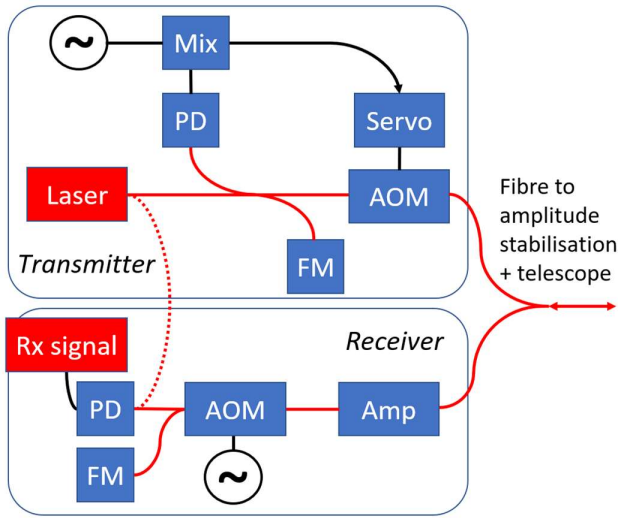

Fig 5: Block diagram of phase stabilisation system. *PD:* photodetector, *FM:* Faraday mirror, *AOM:* acousto-optic modulator. Red denotes an optical signal, while black is electronic.

## 4. Commissioning Status

The telescope is in the process of being relocated from the commissioning lab space to a dome on the roof of the physics building. However, substantial progress has already been made in commissioning various systems.

The telescope is able to track satellite paths by ingesting TLEs, or by using the drone GPS coordinates to generate an artificial TLE on demand. Using this method, the telescope can be slewed to a model satellite mounted on a second floor walkway approximately 20 m from the telescope. This model, like the drone, is equipped with a retroreflector to return the laser signal and beacon LEDs for the machine vision.

Once the beacon LEDs are in the field of view of the machine vision camera, offset rate commands are sent to the telescope to move the target to the "hotspot" location where the returned laser beam will be visible on the QPD. With the machine vision loop closed, the telescope can track the model satellite when carried by hand and moved in random paths. This is more taxing on the system than an operational situation since it is purely the machine vision driving the telescope. In practise, the machine vision will only be needed to make minor corrections to the path provided by the GPS/TLE.

While the machine vision lens and the telescope are co-aligned, they are not co-located and therefore the location of the machine vision hot spot is not dead centre, but rather offset by an amount dependent on the distance to the target. During indoor testing, the hotspot was found by visually aligning the beam with the target, manually aligning the tip/tilt mirror to centre the signal on the QPD, and recording the location of the beacon on the machine vision image. For drone testing, when a misaligned beam will not be visible against a fixed background, a spiral search will be used.

The amplitude-stabilisation loop has been tested with the piezo-actuated TTM, with the stepper TTM still to be incorporated. The TTM controller units interface with the control PC via USB, allowing control from within the custom WAOGS software interface. This also allows access to the QPD error signals, allowing low frequency errors to be offloaded to the telescope to preserve the dynamic range of the tip/tilt system.

Phase-stabilisation has been demonstrated in various point-to-point links, and retroreflected folded links [13, 24]. This system is largely independent of the telescope (functionally and physically), and as such will be essentially "plug and play" with the ground station taking the place of the tip/tilt stabilisation terminals used previously.

## 5. WAOGS Phase II: Path to Operations

To transition from commissioning with a low altitude drone to operations with spacecraft in LEO and beyond, some re-optimisation and upgrades will be needed. The most fundamental of which will be changes to the amplitude-stabilisation optics.

For commissioning, the ground station is configured to transmit a 34 mm beam, which is suitable for the distance to the drone and the size of the retroreflector. For transmitting to space, the full telescope aperture will be needed to minimise divergence of the signal and for collecting as much of the diverged incoming signal as possible. This will require replacing the $f = 300$ mm lens to increase the beam expansion factor to $\sim 300\times$. A single $f = 15$ mm lens would introduce severe spherical aberration, so a series of lenses will be needed to achieve the required beam expansion.

For low altitude drone tests, the beam size is smaller than the Fried Parameter $r_0$ for the level of integrated turbulence along the path. This means that tip/tilt correction is sufficient for amplitude stabilisation. For space-to-ground scenarios where incoming beam and receiver sizes will span many Fried parameters, higher order adaptive optics will be needed for full amplitude-stabilisation. One task of the commissioning phase will be to characterise the atmospheric properties of the site, which will inform how many actuators will be needed. This will also require a commensurate upgrade in the wavefront sensing that will see the QPD replaced with a Shack-Hartmann sensor.

Since spacecraft are in motion, and the speed of light is finite, transmitter pointing must lead the target. The significance of this is that the beacon used for amplitude-stabilisation (whether it be the signal transmitted from the spacecraft or a signal reflected from it) will be offset by the point ahead angle, meaning the transmitted and received beams follow different paths, and experience different turbulence. The higher order adaptive optics corrections will be applicable only to the received signal, and not the transmitted, if the point ahead angle is greater than the isoplanatic angle, the angle at which turbulence through the two paths remains correlated. It is typically on the order of ~10 arcseconds (tens of μrad), which is less than the point ahead angle for most LEO orbits. In this case, only tip/tilt correction remains valid (tip/tilt retains correlation over an angle ~4 times greater than the isoplanatic angle) [26, 27]. The higher order AO will need to be reconfigurable; correcting uplink and downlink when possible and being removed from the uplink path as needed.

Another substantial change will be required for the machine vision. Like GPS, TLEs are only suitable for coarse pointing so machine vision is still required. Unlike the drone scenario, where the short distance makes LEDs readily visible with a small aperture imaging system, spacecraft beacons will be significantly dimmer. The positional uncertainty of TLEs does not warrant the large field of view currently needed for drone acquisition, so the machine vision system can be incorporated into the telescope optics with the increased collecting area greatly improving the sensitivity of the machine vision system. The visible camera currently in use will be replaced with a NIR detector to allow the downlink signal itself to be used as the machine vision beacon.

## 6. Discussion

The centrepiece of WAOGS will be its phase-stabilisation capability. While primarily developed to facilitate extremely stable frequency transfer and timescale comparison of optical atomic clocks over large distances, free space phase-stabilisation can be used to refine frequency mismatch between incoming signal and the local oscillator (LO) caused by uncertainty in Doppler shift of a satellite throughout its orbital pass. As with the telescope tracking, a TLE can generate an estimate of the line-of-sight velocity but some form of closed-loop control is needed to maintain a spatial and temporal lock on the signal. Just as amplitude stabilisation corrects true pointing errors in addition to turbulence beam wander, phase-stabilisation can correct line-of-sight variations from sources other than turbulence.

This sensitivity to variations in Doppler shift can be applied to orbitography; with range rate precisions of ~nm s$^{-1}$, coherent optical Doppler orbitography offers orders-of-magnitude improvement over conventional Satellite Laser Ranging (SLR) or Doppler radar allowing more accurate tracking of spacecraft and space debris, and refinement of the International Terrestrial Reference Frame (IRTF) [24].

Owing to the weak wavelength dependence of the phase noise, the correction applied by the phase-stabilisation servo loop is not limited to the wavelength of measurement but will suppress noise across a range of wavelengths. By combining this technology with wavelength division multiplexing (WDM) and currently demonstrated hardware, ground-to-space links of ~40 Tbps are feasible [21].

Amplitude- and phase-stabilised coherent free space links will also benefit fields beyond communications. The establishment of a global network of atomic clocks, limited only by the clocks themselves, will revolutionise experiments in fundamental physics in areas such as General Relativity [25], variability of fundamental constants [26], and dark matter [27].

## 7. Conclusion

This paper details the architecture of the Western Australian Optical Ground Station and the commissioning undertaken thus far. Each system has been tested individually and as an integrated whole where possible. The telescope tracks spacecraft TLEs, and machine vision is able to maintain the beam retroreflected by a moving target locked on the QPD. Amplitude- and phase-stabilisation have been demonstrated over turbulent horizontal links. Once the telescope is installed with sky access, we expect rapid progress in translating our amplitude- and phase-stabilisation technology to vertical links to an airborne, mobile target.

## Acknowledgments

This project is funded by the Australian Research Council's Centre of Excellence for Engineered Quantum Systems (EQUS, CE170100009) and Goonhilly Earth Station Ltd. Additional resources are provided by the SmartSat Cooperative Research Centre (CRC. Research Project 1-01) and the Australian Space Agency (Moon to Mars Demonstrator Feasibility Project MTMDFG000001). D.G. is supported by a Forrest Research Foundation Fellowship. B.D-M. is supported by an Australian Government Research Training Program (RTP) Scholarship, and B.D-M. and S.K. are supported by SmartSat CRC Scholarships. The WAOGS CDK700 telescope was graciously donated by Colin Eldridge.## References